# Linear stability analysis of hypersonic boundary layers computed by a kinetic approach

## A semi-infinite flat plate at $4.5 \leq M_\infty \leq 9$

**Angelos Klothakis · Helio Quintanilha Jr. · Saurabh S. Sawant · Eftychios Protopapadakis · Vassilis Theofilis · Deborah A. Levin**

Effort sponsored by the Office of Naval Research under grant no. N00014-1-20-2195 "Multi-scale modeling of unsteady shock-boundary layer hypersonic flow instabilities" (Dr Eric Marineau, PO). The US Government is authorized to reproduce and distribute reprints for Governmental purpose not withstanding any copyright notation thereon.

A. Klothakis
Technical University of Crete, Chania, Crete, 73100, Greece, and
University of Liverpool, Liverpool, Brownlow Hill, England L69 3GH, United Kingdom E-mail: aklothakis@gmail.com

H. Quintanilha Jr
University of Liverpool, Liverpool, Brownlow Hill, England L69 3GH, United Kingdom E-mail: helio@liverpool.ac.uk

Eftychios Protopapadakis
National Technical University of Athens, 15780 Athens, Greece

I. Theofilis
University of Liverpool, Liverpool, Brownlow Hill, England L69 3GH, United Kingdom, and
Escola Politecnica, Av. Professor Mello Moraes 2373, Universidade São Paulo, Brasil

S.S Sawant · D.A. Levin
University of Illinois at Urbana-Champaign, Urbana, Illinois 61801, USA

**Abstract** Linear stability analysis is performed using a combination of two-dimensional Direct Simulation Monte Carlo (DSMC) [3] methods for the computation of the basic state and solution of the pertinent eigenvalue problem, as applied to the canonical boundary layer on a semi-infinite flat plate. Three different gases are monitored, namely nitrogen, argon and air, the latter as a mixture of 79% $N_2$ and 21% $O_2$ at a range of free-stream Mach numbers corresponding to flight at an altitude of ~55km. A neural network has been utilised to predict and smooth the raw DSMC data; the steady laminar profiles obtained are in very good agreement with those computed by (self-similar) boundary layer theory, under isothermal or adiabatic wall conditions, subject to the appropriate slip corrections computed in the DSMC method [1,2].

The leading eigenmode results pertaining to the unsmoothed DSMC profiles are compared against those of the classic boundary layer theory [32]. Small quantitative, but no significant qualitative differences between the results of the two classes of steady base flows have been found at all parameters examined. The frequencies of the leading eigenmodes at all conditions examined are practically identical, while perturbations corresponding to the DSMC profiles are found to be systematically more damped than their counterparts arising in the boundary layer at the conditions examined, when the correct velocity slip and temperature jump boundary conditions are imposed in the base flow profiles; by contrast, when the classic no-slip boundary conditions are used, less damped/more unstable profiles are obtained, which would lead the flow to earlier transition. On the other hand, the DSMC profiles smoothed by the neural network are marginally more stable than their unsmoothed counterparts.

A vortex generator (VG) introduced into the boundary layer downstream of the leading edge and pulsed at rather large momentum coefficient, $C_\mu = 0.27$, and scaled frequency $F^+ \approx 0.98$ [17], is used to generate linear perturbations that decay along the plate, as expected from the low value of the Reynolds number, $Re_\delta = 290$, in this numerical experiment. The damping rate diminishes monotonically as the VG is placed at successive downstream positions along the plate. The characteristics of the oscillation generated in the boundary layer are predicted accurately by linear stability analysis of the undisturbed profile at the location of VG placement. Most interestingly, the effect of the generated perturbation is felt well outside of the boundary layer, generating oscillations of the leading edge shock that synchronise with linear perturbations inside the boundary layer.

# 1 Introduction

In recent decades considerable efforts have been devoted to the design of high-velocity flying vehicles, sub-orbital vehicles and microelectromechanical systems [11, 34, 12] in all of which rarefied gas flows are encountered [22, 5, 30]. Depending on the exact value of the Knudsen number, $Kn$, modifications to the boundary conditions used in the compressible Navier-Stokes equations are required, for the latter to properly capture flow physics. Flows in which $0.01 \lesssim Kn \lesssim 0.1$ belong to the slip regime and need require appropriate treatment of the wall boundary conditions to account for velocity slip and temperature jump[63]. At $Kn > 0.1$ alternative methodologies based on kinetic theory, such as DSMC [3], moment equations [18] or numerical solutions of the full Boltzmann equation [9], are required. Although kinetic theory methods can be applied to all flow regimes [63, 9], their use in continuum and slip flow regimes is computationally very intensive and methods based on the Navier-Stokes equations are preferable. However, a main drawback of PDE-based description of compressible flows remains its inadequate description of the shock layer structure, as already discussed by Liepmann *et al.* [29].

Renewed interest in maneuverable sustained hypersonic flight at altitudes around 50 $km$ inside the earth's atmosphere brings to the fore the question of laminar-turbulent transition prediction on any part of the vehicle surface, and especially on axisymmetric vehicle forebodies and on lifting surfaces, typically modeled as circular-base cones and flat plates, respectively. Intense experimental and numerical efforts are underway [6, 53], mostly employing classic boundary layer linear stability theory [31–33, 10] to predict and control linear instability mechanisms leading boundary layer flow to transition and turbulence and prevent a multifold increase in the thermal protection requirements and decrease of vehicle range. Such linear stability analysis approaches either altogether exclude the shock from the analysis, or include it in the underlying (steady) base flow by appropriate modifications of the boundary conditions used at the boundary-layer edge [55, 35]. In doing so, the internal shock layer structure (which is inaccessible to the Navier-Stokes equations) and its potential effect on boundary layer stability is either neglected or modeled through the boundary conditions. It would thus appear natural to apply kinetic theory methods to address linear flow instability; however, to-date it has not been demonstrated that kinetic theory methods can meet the long known stringent requirements on the quality of the base flow (and its first and second derivatives) in order for reliable stability analysis results to be obtained [33]; examination of this issue is the first motivation of the present contribution.

Recent results of application of global linear stability theory concepts to DSMC simulation results are encouraging. In a series of papers Tumuklu *et al.* [59, 61] demonstrated strong coupling between the shock structure and linear instability of a laminar two-dimensional separation bubble in Mach 16 axisymmetric flows over a $25°-55°$ double-cone, as well as in Mach 7 flows over a $30°$-$55°$ double-wedge configuration [60]. On double-cones, the presence of $\lambda$-shocks and oscillations of the detached and separation shocks at a Strouhal number, $St$, of 0.078 has been documented [61]. Subsequent extension of the analysis to 3-D, spanwise-homogeneous flow over the same double-wedge configuration by Sawant *et al.* [52, 51] showed the presence of linearly growing, self-excited, small-amplitude, 3-D perturbations inside the separation bubble as well as in the interior of separation and detached shock layers, which, in turn, leads to low-frequency unsteadiness of the triple point at $St \sim 0.02$. The lat-ter authors have also highlighted the importance of modeling the well-known, bimodal internal structure of shocks in hypersonic flows, where the internal nonequilibrium zone is shown to exhibit two-orders of mangitude larger amplitude and lower frequency broadband, having mean $St$ 0.01, than the equilibrium zone of freestream [49, 50].

This paper addresses the ability of DSMC to generate steady and unsteady laminar boundary layer flows of sufficient quality for linear stability analysis to be performed. Compressible laminar boundary layers developing on a semi-infinite flat plate have been computed by Gallis and Torczynski [14] using the BGK model and by Kumar *et al.*[28] using both BGK and DSMC simulations. Here, DSMC is employed to compute flows over a range of Mach numbers, using three gases with distinct thermodynamic properties. The base flows are obtained using the Open Source DSMC code SPARTA of the Sandia National Laboratory [15]. DSMC results on the wall-normal velocity and temperature gradients were used to formulate the velocity slip and temperature jump boundary conditions as described by the Maxwell/von Smoluchowski theory [37, 54] and later revised by Beskok *et al.* [2]. The DSMC data, smoothed using a neural network, are compared with those delivered by the boundary layer theory; such comparisons are permissible since the flows addressed are in the slip regime. Subsequently, both sets of profiles are analysed with respect to their linear stability and the eigenspectra obtained are compared. Finally, a periodically pulsating jet is used as a vortex generating device in the DSMC to inject particles into the boundary layer and analyse the characteristics of the oscillations observed in the boundary- and the shock layer. The frequency and damping rate of these oscillations are compared with the respective eigenvalues of the leading linear stability theory boundary layer perturbations. The paper is organised as follows: Section 2 presents the essential implementation details of the DSMC simulations performed, as well of the smoothing approach employed. Section 3 presents in some detail the boundary layer equations and boundary conditions used to obtain comparison profiles; some details of the classic linear stability theory problem solved, including the scales used to convert between dimensional and dimensionless results and enable comparisons close this section. In Section 4 the results obtained are presented; first, the comparisons of the steady laminar base states in two of the three cases analysed are shown, followed by comparison of the eigenvalue spectra and amplitude functions pertaining to the DSMC and the boundary layer profiles. This section closes with the description of the introduction of boundary layer perturbations and the evolution and correlation of the generated instability waves inside the boundary layer and along the shock layer. Concluding comments are offered in Section 5.

## 2 The Direct Simulation Monte Carlo method

The Direct Simulation Monte Carlo (DSMC) method [4] is well established method of choice to simulate rarefied gas flows. Its implementation is based on the ability to simulate a large (and ever increasing, commensurate with increasing computing capabilities) number of 'simulator' particles, each of which represents a large (and ever decreasing, commensurate with increasing computing capabilities) number of real gas molecules. The computational domain is divided in cells that define geometric boundaries and computational volumes, in which macroscopic flow parameters are computed by simulating motion and collisions of the aforementioned simulator particles [3]. The DSMC is an iterative algorithm and can be decomposed in four steps. First, particles are moved to new positions during a time step. Second, indices are assigned to the particles moved in new cells. Third, collision pairs are selected and collisions are performed. Fourth, macroscopic flow properties are calculated by averaging temporally the microscopic properties of the flow and the procedure is restarted. By its very nature, DSMC lends itself to massive parallelization and the method, use of which is essential in the rarefied gas regime, is also gaining traction recently in performing slip (and even continuum) regime computations as computing hardware capabilities increase. Algorithmic and implementation details in the standard references of Bird [3, 4].

2.1 DSMC Implementation

DSMC simulations performed herein have used the massively parallel open-source code SPARTA[15, 41] developed in Sandia National Laboratories, validations of which against several benchmark test cases have been presented elsewhere [24]. Computing resources for the present high-resolution simulations were made available on the Archer and Archer-2 UK supercomputing resources. In all of the cases run in this work the Variable Hard Sphere (VHS) model was used for all three gases analysed, argon, nitrogen and air as a mixture of nitrogen and oxygen. For the accurate modeling of particle collisions the near collision partner algorithm was used and collision partners were selected from within the distance the particle travels in one timestep [13]. Both adiabatic and isothermal interactions of the particles with the walls have been addressed. In the simulations that follow, air, argon and nitrogen have been used. In the case of air the Maxwell model was used with coefficients governing surface accommodation set to zero. For argon, where no rotational degrees of freedom exist, the pre-collision and post-collision translational energies of particles at the wall is kept constant. Physical parameters of the gases and conditions simulated are shown in table 1, with further details on the DSMC parameters shown in table 2. A typical result for the streamwise velocity component of the steady state obtained in argon is shown in the upper part of figure 1. The lower part of the same figure shows the evolution of pressure at the edge of the boundary layer, $y = 0.0014m$, along the streamwise coordinate, $x$, over the entire length of the plate for both air and argon DSMC simulations. The effect of the evident small acceleration of the flow on the stability of the boundary layer will be discussed in section 4.

## 2.2 DSMC Data Smoothing

In statistical flow simulations, fluctuations in the flow parameters over a mean arise at each time step, making calculation of the derivatives of macroscopic flow parameters a rather challenging task. On the other hand, well-resolved and accurate base flow quantities and their spatial derivatives are essential for the calculation of reliable linear stability results. Smoothing of the DSMC data is thus required and it is well-known [48, 21] that choice of an inappropriate smoothing method can cause severe changes to the original data and can lead to incorrect interpretations of the results. In our past work different smoothing techniques have been utilised, e.g. the WPOD method discussed by Tumuklu *et al.*[60]. Here we employ a stack autoencoder neural network, which we use in an unsupervised way in order to smooth the signal from the DSMC simulations of Cases 1 and 2. Deep learning neural networks such as this are widely used in several different fields such as computer vision, speech recognition and human tracking [19, 16]; some technical details of the present implementation follow.

An autoencoder is a deep learning neural network that is constructed in a way that it is able to copy its input to its output. In contrast with other neural networks only one hidden layer is used to represent the input data, such that the autoencoder network be viewed in two parts: the first is an encoder function $h^* = f(x)$, which is decoded in the second part, where the reconstruction of the input data is calculated as $r^* = g(h^*)$. Autoencoders can also work with incomplete/undercomplete and sparse data, such as the undersampled DSMC simulation profiles. An undercomplete set is trained such as the input copying task results in the hidden layer capturing useful properties by constraining it to have a smaller dimension than the input. A network that has learnt to represent undercomplete data is able to capture the most important underlying information; in the case of the DSMC simulation profiles, this information is the mean value of the quantities of interest, in which the noise level has been reduced. The learning procedure can be described as minimizing a loss function $L(x, g(f(x)))$, where the $L$ is a loss function penalizing the $g(f(x))$ or dissimilar to $x$; as an example an $L$ function can be the mean-squared error. If the decoder is linear and $L$ is the mean squared error then the undercomplete autoencoder learns the principal subspace of the training data as a side effect.

In the present work the neural network is trained using four wall-normal velocity profiles, taken at locations $x = 0.2, 0.3, 0.4$ and $0.5$ m from the plate leading edge, and a total of 1200 data points in each training set. After the network was trained, the smoothed wall-normal velocity profile at a single, randomly chosen location $x = 0.7$ m, was compared with simulation data to verify the quality of the neural network training; the result obtained for air at the parameters shown in tables 1 and 2 is shown in figure 3.

## 3 The Compressible Laminar Boundary Layer

For completeness, the equations governing the steady laminar two-dimensional compressible boundary layer, the boundary conditions used and the linear stability equations employed are summarised here.

### 3.1 Governing equations and boundary conditions

The boundary layer equations on a Cartesian frame of reference read

$$\frac{\partial(\rho^* u^*)}{\partial x^*} + \frac{\partial(\rho^* v^*)}{\partial y^*} = 0, \tag{1}$$

$$\rho^* u^* \frac{\partial u^*}{\partial x^*} + \rho^* v^* \frac{\partial u^*}{\partial y^*} = \rho_e^* u_e^* \frac{du_e^*}{dx^*} + \frac{\partial}{\partial y^*}\left(\mu^* \frac{\partial u^*}{\partial y^*}\right), \tag{2}$$

$$\rho^* u^* \frac{\partial h^*}{\partial x^*} + \rho^* v^* \frac{\partial h^*}{\partial y^*} = \rho_e^* u_e^* \frac{dh_e^*}{dx^*} + \frac{\partial}{\partial y^*}\left(k^* \frac{\partial T^*}{\partial y^*}\right) + \mu^* \left(\frac{\partial u^*}{\partial y^*}\right)^2, \tag{3}$$

an asterisk denoting dimensional quantities. The velocity along the streamwise and wall-normal direction are denoted by $u^*$ and $v^*$, respectively, $\rho^*$ is the flow density, $T^*$ is temperature and the subscript $e$ denotes conditions at the edge of the boundary layer (as opposed to the subscript $\infty$, which is reserved for (pre-shock) free-stream conditions [20,31,7,62].

At the values of the Knudsen number examined slip flow exists and the above equations require to be closed by wall boundary conditions that account for the velocity slip and temperature jump. The velocity slip and temperature jump boundary conditions used in this work are those of Von Smoluchowski [54] with a correction to the Maxwell conditions [37] as improved by Beskok *et al.*[2], who modified these boundary conditions to include higher order terms,

$$u^*_{\text{slip}} = \frac{1}{2}\left[\frac{u^*_\lambda + (1-\sigma_v)u^*_\lambda + \sigma_v u^*_{\text{wall}}}{\lambda}\right] + \frac{3\sigma_v}{8}\frac{\mu^*}{\rho^* T^*}\frac{\partial T^*}{\partial x^*}, \tag{4}$$

$$T^*_{\text{slip}} = \frac{\frac{2-\sigma_T}{Pr}\frac{\gamma}{\gamma+1}\frac{T^*_\lambda + \sigma_T T^*_{\text{wall}}}{\lambda}}{\sigma_T + \frac{2-\sigma_T}{Pr}}. \tag{5}$$

Here $u^*_\lambda$ and $T^*_\lambda$ denote the values of the streamwise velocity and temperature respectively at a distance of one mean free path, $\lambda$, from the wall, while $u^*_{\text{wall}}$ is the velocity of a moving wall and $T^*_{\text{wall}}$ is the specified wall temperature, when isothermal simulations are performed. This form of the equations as reported by the authors allows these boundary conditions to obtain numerical solutions of the Navier-Stokes equations in the slip regime and up to very high Knudsen numbers, $Kn \leq 0.5$ [2]. The accommodation coefficients for velocity and temperature [5, 1, 27] are denoted by $\sigma_v$ and $\sigma_T$, respectively, and are both taken equal to unity. The wall values delivered by equations (4-5) have been found to be in agreement with the slip velocities and temperature jumps computed in the DSMC, the latter shown in table 3.

## 3.2 Scaling of the boundary layer equations

In order to perform stability analysis all the dimensional components in equations 1 - 4 are recast in non-dimensional form. The velocities $u^*$, $v^*$, density $\rho^*$, dynamic viscosity $\mu^*$, and temperature, $T^*$ are scaled by their respective boundary layer edge values to become:

$$u = u^*/u_e, \ v = v^*/u_e, \ \rho = \rho^*/\rho_e, \ \mu = \mu^*/\mu_e, \ T = T^*/T_e. \tag{6}$$

Spatial coordinates are scaled as proposed by Mack [31], introducing

$$\xi^* = x^* \quad \text{and} \quad \eta = \frac{y^*}{x^*}\sqrt{Re_x}, \tag{7}$$

with

$$Re_x = \frac{\rho_e u_e x^*}{\mu_e}. \tag{8}$$

The Reynolds number used in the linear stability analysis then becomes

$$Re = \sqrt{Re_x}. \tag{9}$$

Substitution into the equations (1 - 3) yields the system solved for a zero-pressure-gradient compressible boundary layer:

$$\rho v - \frac{1}{\sqrt{Re_x}}\left(\eta g - g\right) = 0, \tag{10}$$

$$\frac{d}{d\eta}\left(\mu \frac{du}{d\eta}\right) + g\frac{du}{d\eta} = 0, \tag{11}$$

$$\frac{d}{d\eta}\left(\frac{\mu}{Pr}\frac{d\theta}{d\eta}\right) + g\frac{d\theta}{d\eta} + 2\mu\left(\frac{du}{d\eta}\right)^2 = 0, \tag{12}$$

where

$$g = \frac{dg}{d\eta} \equiv \frac{1}{2}\rho u, \tag{13}$$

$$\theta = \frac{T^* - T_e^*}{T_0^* - T_e^*}, \tag{14}$$

$$T_0^* = T_e^* + \frac{u_e^2}{2c_p^*}, \tag{15}$$

$$\frac{T^*}{T_e} \equiv T = 1 + \frac{\gamma - 1}{2}M_e^2 \theta, \tag{16}$$

and $Pr$ is the Prandtl number defined as:

$$Pr = \frac{\mu_e c_p}{\kappa_e}. \tag{17}$$

The system of equations is closed by considering the dependence of viscosity on temperature according to

$$\mu = \mu_{\text{ref}}\left(\frac{T}{T_{\text{ref}}}\right)^\omega, \tag{18}$$

where $\mu_{\text{ref}}$ and $T_{\text{ref}}$ are reference viscosity and temperatures, the gas constant $c^*$ is defined as $c_e = \frac{\gamma R}{\gamma-1}$, $V_r = c^*/\dot{c}$ and $\varpi$ is the viscosity exponent. Reference values for these constants for the gases considered herein, provided by Bird [3], are shown in tables 1 and 2. The boundary layer system of equations can be solved subject to the boundary conditions

$$u(\eta = 0) = u_{\text{slip}}/u_e \tag{19}$$

$$u(\eta \to \infty) = 1 \tag{20}$$

$$T'(\eta = 0) = 0 \quad \text{or} \quad T(\eta = 0) = T_w \tag{21}$$

$$T(\eta \to \infty) = 1. \tag{22}$$

Results obtained at $u_{\text{slip}} = 0$ have been validated by comparison against standard references [43, 38]. The wall-slip and temperature jump values arising in the present simulations are shown in table 3.

### 3.3 Linear Stability Analysis

Flow stability analysis is based on the compressible Navier-Stokes and continuity equations, in dimensionless form:

$$\frac{\partial \rho}{\partial t} + \nabla \cdot (\rho \mathbf{V}) = 0, \tag{23}$$

$$\frac{\partial (\rho \mathbf{V})}{\partial t} + \nabla \cdot (\rho \mathbf{V}\mathbf{V}) = -\nabla p + \frac{1}{Re} \nabla \cdot \sigma \tag{24}$$

$$\rho \left[ \frac{\partial T}{\partial t} + (\mathbf{V}.\nabla) T \right] - Ec \left[ \frac{\partial p}{\partial t} + (\mathbf{V}.\nabla) p \right] = \frac{1}{RePr} \nabla \cdot (k \nabla T) + \frac{Ec}{Re} \frac{1}{2} (\nabla \mathbf{V} + \nabla \mathbf{V}) : \sigma \tag{25}$$

$$p = \frac{\rho T}{\gamma M^2} \tag{26}$$

where,

$$\sigma = [\mu (\nabla \mathbf{V} + (\nabla \mathbf{V})^T)] + \nabla [\mu_2 (\nabla \cdot \mathbf{V})]$$

is the viscous stress tensor, $\mathbf{V} = \{u,v,w\}^T$ is the velocity vector, $\rho$ is the density, $p$ is the pressure, $T$ is the temperature, $\mu$ is the first coefficient of viscosity, $\mu_2$ is the second coefficient of viscosity and $k$ is the thermal conductivity. The dimensionless parameters are the Mach number $M$, the Reynolds number $Re$, the Prandtl number $Pr$ and the Eckert number $Ec$.

The evolution in space and time of small amplitude perturbations imposed upon a base flow is very well described by the Linearized Navier-Stokes Equations (LNSE). The linearization is made from the decomposition of the state vector $\mathbf{q} = (\rho, u, v, w, T)$ into a steady laminar base flow and a small amplitude disturbance such as,

$$\mathbf{q}(\mathbf{x}, t) = \bar{\mathbf{q}}(\mathbf{x}) + E\tilde{\mathbf{q}}(\mathbf{x}, t), \qquad with \quad E \ll 1. \tag{27}$$

The $O(1)$ equations resulting from the above are those which govern the base state and are satisfied by construction, the $O(E^2)$ equations are neglected, on account of smallness of the perturbation amplitudes, while the $O(E)$ equations are the LNSE, that need to be solved, either as an eigenvalue problem (modal analysis) or as an initial value problem (non-modal analysis). In equation (27) the $\mathbf{x}$ is the space vector coordinator while $t$ is time and $E$ is a very small amplitude disturbance.

The decomposition (27) is valid for one-, two- and three-dimensional base flows. In this work local analysis is performed, in which only the wall-normal spatial direction is taken as inhomogeneous in both the base flow and the amplitude functions. Two-dimensional parallel flow is assumed and only the streamwise velocity and temperature components extracted from the DSMC simulation are utilised. After performing separation of variables, the two homogeneous directions are decomposed in Fourier space and the linearized equations of motion are re-written as a system of ODE equations extensively discussed by Mack [31–33]. Assuming $x$ and $z$ as the homogeneous spatial directions, such that the base flow dependents only on the $y$ spatial coordinate in a Cartesian frame of reference, modal perturbations can be written in a local context as:

$$\tilde{\mathbf{q}}(x, y, z, t) = \hat{\mathbf{q}}(y) \exp^{i(\alpha x + \beta z - \omega t)}. \tag{28}$$

In a temporal framework, $\alpha$ and $\beta$ are the wavenumbers in the streamwise and spanwise directions, respectively and $\omega$ is a complex eigenvalue solved for. Its real part, $\omega_r$, is related to the perturbation angular frequency, while the imaginary part, $\omega_i$, represents the damping rate. Dimensional quantities are made dimensionless using an appropriate length scale, $L_{\text{ref}}$ and the edge streamwise velocity component, $u_e$, as will be discussed shortly.

Substituting equation (28) into the compressible LNSE the one-dimensional eigenvalue problem introduced and solved by [32, 33] is obtained. Written in compact form this reads

$$\mathbf{L}\hat{\mathbf{q}} = \omega\hat{\mathbf{q}}, \tag{29}$$

where $\hat{\mathbf{q}}$ is an one-dimensional amplitude function. The solution of this eigenvalue problem determines if the flow is stable or unstable in a modal stability context [57]. This depends on the sign of the imaginary part of the complex parameter $\omega$. If $\omega_i < 0$ then the flow is modally stable and the perturbations decay in time. Otherwise, if $\omega_i > 0$ the flow is modally unstable and the perturbations grow exponentially. The eigenvalue problem (29) is solved using the *Li* near *G*lobal instability for *H*ypersonic *T*ransition (*LiGHT*) code [45, 46, 58, 44]. The in-house code is written in Fortran and it is suite of subroutines for the massively parallel solution of complex non-symmetric eigenvalue problems (EVP) and Singular Value Decomposition (SVD) problems arising in linear fluid flow instability. In this paper, the local stability version of the code is used.

*3.3.1 Spatial discretization*

The wall-normal direction $\eta$, introduced in (7) is discretized using high-order spectral collocation based on Chebyshev Gauss-Lobatto (CGL) points $x_j^*$,

$$x_j^* = \cos\frac{j\pi}{N}, \quad j = 0, N. \tag{30}$$

The solution of boundary-layer problem requires clustering of points close to the wall. Therefore, the well-tested mapping [56, 39, 46] is used:

$$\eta_j = l\frac{1 - x_j^*}{1 + s + x_j^*}, \tag{31}$$

where $s = 2l$ and $l = \eta_{\frac{1}{2}}/(1 - 2\eta_{\frac{1}{2}})$. In boundary layer linear stability problems half of the collocation points are placed between the wall and the parameter $\eta_{\frac{1}{2}}$ [36]; here $\eta_{\frac{1}{2}} = 0.2$ has been used.

## 4 Results

4.1 Gas conditions and geometric parameters

Three sets of cases have been investigated: in Case 1 (air) and Case 2 (argon) the same free-stream conditions, $T_\infty = 190\ K$, $P_\infty = 34.04\ Pa$, have been used, corresponding to flight at an altitude of $\sim 55\ km$. In Case 3 (nitrogen) a set of parameters was chosen in order to obtain Mach 9 flow in the free-stream. The details of the simulation conditions are summarized in 1. For the first two cases these specific parameters were selected in order to try to obtain a $M\ 4.5$ at the edge of the boundary layer and compare results in relation to Mack's second mode. Air was simulated as a gas mixture of 79% Nitrogen ($N_2$) and 21% Oxygen ($O_2$). Due to the relatively low Mach number no chemical reactions between the $N_2$ and $O_2$ have been considered.

The full set of parameters used in the simulations can be found in table 1. In all three cases the flat plate upon which the boundary layer develops has a thickness of $d = 0.5\ mm$ and a circular nose with radius $r = 0.5 mm$; in the first two cases the plate has a length of $L = 1\ m$, while in the third case the plate length is shortened to $L = 0.15\ m$, in order to keep the cost of unsteady DSMC computations that will be described in 4.3 within reasonable limits. In order for the domain to accommodate all flow effects expected a height of $0.4m$ was used in Cases 1 and 2, while a domain height of $0.05m$ was used in Case 3. Two-dimensional DSMC simulations were performed, the parameters of which are shown in table 2. The east side of the domain is defined as inflow, where particles enter the simulation domain and the west side as outflow, where every particle crossing the boundary is deleted. North and south sides are defined as symmetry planes, where every particle that hits these boundaries is specularly reflected back in the simulation domain. In the third case the surface boundary condition is set to diffuse reflection with full surface accommodation.

In all three cases in order to assure accuracy of particle collisions the collision partners were selected within a sphere with radius equal to the distance that the particle travels in one timestep according to the near-neighbor algorithm [13]. On the flat plate surface either isothermal or adiabatic diffuse reflection boundary condition has been imposed. Isothermal results were reported in [25, 26] and here mostly adiabatic calculations are shown. Finally, in order to ensure numerical (DSMC) as well as physical (spatial and temporal) accuracy the cell size in each of the domains employed has been kept below one mean free path $\lambda$, while the time-step employed led to every particle entering a cell required on average five time steps to cross the cell.

Figure 4 shows a comparison of velocity and temperature profiles for argon and air at a given location $x = 0.7\ m$ along the plate, where $\sqrt{Re_x} \approx 200$, as obtained by DSMC and the compressible similar zero pressure gradient boundary layer equations. The different properties of the gases result in substantially different boundary layer velocity and (especially) temperature profiles, the wall temperature for $Ar$ being 75% higher than that of air. When comparing the DSMC and boundary layer predictions it can be seen that the former is marginally thicker than the latter. The quantitative differences observed are attributed to the mild pressure gradient of the DSMC results at the streamwise location monitored, as well as to the slightly different value of the Prandtl number, which is taken to be constant in the boundary layer approximation. Analogously good quantitative comparisons of base flow profiles obtained by DSMC under the isothermal wall boundary condition and compressible boundary layer solutions subject to slip velocity and temperature jump boundary conditions have been discussed recently in [25].

## 4.2 Linear stability analysis of steady boundary layer profiles obtained by DSMC

Eigenspectra pertaining to the DMSC- and boundary-layer profiles obtained for air in Case 1 and Argon in Case 2 at a given two-dimensional wavenumber $a = 0.2$ are respectively shown in figures 5 and 6; the corresponding amplitude functions can be found in figures 7 and 8. In the plots showing the amplitude functions of air the location of the generalised inflection point (GIP) is also indicated by a dashed horizontal line. Both sets of results show that the linear instability properties of the profiles extracted from the DSMC simulations and those computed in the corresponding boundary layer approximation are qualitatively identical and actually in close quantitative agreement. Only damped eigenvalues have been found at all sets of parameters, as expected from the relatively low values of $Re = \sqrt{Re_x}$.

Interestingly, figure 5 also includes the eigenspectra of compressible boundary layer profiles in which the classic no-slip boundary condition is imposed, as well as that calculated from the neural network smoothing. The least damped discrete mode corresponding to the no-slip boundary layer profile is *less stable* than that pertaining to the profile in which the slip boundary conditions (4-5) have been imposed. The implication is that imposition of the no-slip boundary condition in the base flow profile leads to theoretical prediction of *earlier boundary layer transition*. On the other hand, the least damped discrete eigenvalue pertaining to the smoothed data is marginally more stable than that of the raw DSMC profiles.

The acoustic branches pertaining to all four base flow profiles are practically identical, as are (rather surprisingly), the frequencies of the leading discrete modes of the raw DSMC profile and that of the boundary layer subject to boundary conditions (4-5). The quantitative differences in the damping rates between the air and argon damping rate results may be attributed to the nonzero pressure gradient present in the DSMC results, seen in figure 1 (lower); consequently, DSMC profiles are found to be stronger damped in their boundary layer counterparts.

The results shown are representative of others obtained at different combinations of the (*Re, α, β*) parameters and not shown here for brevity. To the best of the authors' knowledge these results establish, for the first time, the ability of the kinetic-theory based DSMC approach to predict steady laminar base flows of sufficient quality to be in close agreement with the established, Navier-Stokes/boundary-layer-equations-based linear stability theory [31–33]. Our results also demonstrate that imposition of the no-slip boundary condition on the base flow profiles leads to prediction of less damped / more unstable leading discrete eigenmodes compared to those found when the slip boundary condition is imposed.

### 4.3 Vortex Generator definition and evolution of waves introduced in the boundary layer

In the presently studied Case 3, vortical perturbations are introduced inside the boundary layer by a jet of high-density particles through an orifice of diameter $D_o = 1\ mm$, placed at $x_o = 1\ cm$ from the leading edge. The jet pulsates periodically at an excitation frequency $f_e = 277.77\ kHz$ and particles are injected with a velocity magnitude of $U_o = 4000\ m/s$ at an angle of 60 degrees to the plate, at a temperature of $T_o = 245.45\ K$. Each time the jet fires about $10^5$ particles enter the domain and the mass flow rate that passes from a surface with a length of 20 computational cells is $\dot{m} = 3.76 \times 10^{-4}\ kg/s$. The reduced momentum coefficient, $C_\mu$, and reduced frequency, $F^+$, corresponding to the jet actuation are respectively defined [17] as

$$C_\mu = \frac{\rho_o U_o^2 D_o}{\rho_\infty U_\infty^2 L_x} = 0.268, \tag{32}$$

and

$$F^+ = \frac{f_e x_o}{U_e} = 0.983. \tag{33}$$

Here, $U_e$ is the local edge velocity at the location $x_o$ of the orifice. On account of the rather large jet velocity, $U_o$, chosen in the simulations, the reduced momentum coefficient is substantially larger than that which was found to be adequate for the excitation of linear perturbations in the boundary layer experiments of Greenblatt and Wygnanski [17], while the reduced frequency value is of the same order of magnitude as that found in experiments. A full discussion of technical details regarding the implementation of the jet generation can be found in [40]; here the essential details are outlined. At the start of the simulation, properties of the particle group that is going to be inserted such as number density, velocity components and thermal temperature are defined. At every timestep inside each grid cell that overlaps with one or more emitting surface elements the number of particles to be added is computed, according to eq. (4.22) in Bird [3]. The number of particles to be inserted is based on the molecular flux calculated by this equation and also from properties like the ratio $fnum$ of the injected particles to those in the flow, the overall number density, velocity components and temperature of the injected particles, as well as the fraction of the surface element that overlaps with the corresponding insertion grid cell and the cell orientation with respect to the streaming velocity. All of these properties can be user-defined, as detailed in [40].

A qualitative image of the perturbations generated can be seen in figure 9. In the lower image of this figure two white lines are marked, one inside in the boundary layer at a distance $0.001 m$ from the plate surface, and another one close to the shock, at a distance $4.4\ mm$ from the flat plate and at an angle $\varphi = 13.19°$ to the plate. On these lines unsteady flow data are extracted and post-processed; results will be discussed shortly. In figure 10 (upper) the evolution in time of the total number of particles in the simulation is shown. Immediately after the jet firing has started at $t = 0$, an increase in the number of particles can be seen, as expected due to the addition of new particles in the simulation domain and a steady state is reached after $t \approx 0.2\ ms$. As the injection of particles continues, a new periodic state is arrived at around $t \approx 0.54\ ms$, as shown in the inset figure. Results discussed in what follows are extracted from the DSMC simulation during this time window, in which a zero-net-mass state, in which the number of particles entering and exiting the simulation domain is the same, has been established. At around $t \approx 0.67\ ms$ the injection of particles ceases and the flow returns to its previously established steady state, as evidenced, for example, by the streamwise velocity profile at $x = 0.05\ m$, shown before ('pre') and after ('post') the action of the vortex generator in the lower part of figure 10. The entire evolution can be seen in a video submitted as Supplemental Material to this article.

Figure 11 presents the wall-normal component of the velocity as a function of the streamwise spatial coordinate at a height of 1.5 mm, at two successive timesteps, $t_1 = 0.549$ and $t_2 = 0.551\ ms$, within the time interval $0.54 \leq t \leq 0.66$ shortly after the zero-net-mass injection conditions shown in figure 10 had been established. This quasi-periodic signal, as well as others extracted at different heights within the boundary layer, but not shown here for brevity, can be seen to decay exponentially with $x$ along the downstream spatial direction. In order to compare the characteristics of this signal with the theoretical result of the least damped linear instability, an exponential curve fit was constructed, in the form

$$y = Ae^{\sigma x} + C \tag{34}$$

the parameters taking the values $A = 508.86, \sigma = -40.32$ and $C = 25.60$; the curves resulting from damping rates with an error bar of $\pm 10\%$ are also indicated on this figure.

The steady laminar boundary layer profile at $x = 0.04\ m$ was subsequently analysed at the parameters shown in Table 3. Figures 12 and 13 respectively present the eigenspectrum and the eigenfunctions of the leading damped modes of this flow, as predicted by solution of the eigenvalue problem pertaining to the DSMC profile; the respective dimensionless and dimensional results are shown in table 4. Figure 13 presents the amplitude functions of the least damped flow perturbation, denoted as Mode 1, alongside the second in significance linear perturbation at these parameters, denoted as Mode 2 and identified as the compressible analog of a Tollmien-Schlichting (TS) wave. Interesting as it may be, the fact that a TS wave exists in the eigenspectrum of the steady laminar DSMC profile is less significant than the ability of the kinetic theory to capture as damped periodic oscillation the leading eigenmode of the underlying steady boundary layer profile. The dimensional value of the damping rate obtained in the stability analysis is used as $\sigma$ to construct the curve-fit equation (34) plotted in figure 11, where an error bar corresponding to $\pm 10\%$ deviation from the value of $\sigma$ used is also indicated. The results leave little doubt that the damped oscillation generated by the action of the pulsating jet is that captured as Mode 1 in the linear stability analysis of the steady laminar DSMC profile.

Finally, figure 14 presents the wall-normal velocity component inside the boundary layer, already shown in figure 11, alongside the value of the same quantity along the oblique line shown in figure 9 below the shock. It can clearly be seen that the oscillation inside the boundary layer is synchronized with the footprint of the shock oscillation. In order to quantify the oscillations, the two signals are Fourier transformed and the result is shown in the lower part of the figure 14. It can be seen that the wavenumber of the oscillation inside the boundary layer,

$$\frac{1}{L_x} = \frac{a}{2\pi L_{\text{ref}}} \approx 100 \frac{1}{m}, \tag{35}$$

can also be found in the signal measured below the shock, although the peak wavenumber of the shock oscillation is slightly displaced toward lower values; in other words, the waves propagating along the shock are found to have slightly larger wavelengths than those inside the boundary layer. Additionally, it is interesting to note that the lower wavenumber content of the shock oscillation is consistently higher than that inside the boundary layer. The dimensional frequency of the leading damped mode inside the boundary layer can be computed from the (dimensionless) result for $\omega_r$ by

$$f = \omega_r \frac{u_e}{2\pi L_{\text{ref}}} \approx 288.6\ kHz, \tag{36}$$

and is seen to be within 4% of the jet oscillation frequency.

## 5 Summary

Laminar hypersonic flows of air, argon and nitrogen over a flat plate have been computed by highly-resolved DSMC simulations in the Mach number range $4.5 \leq Ma \leq 9$. Boundary layer profiles extracted from the DSMC simulations have been analysed with respect to their linear modal instability. Results in air and argon have been compared with those pertinent to the corresponding compressible zero pressure gradient boundary layer profiles, in which wall slip was taken into consideration. In both gases the acoustic branches obtained in the DSMC and the boundary layer analyses were indistinguishable, while the leading members of the discrete spectrum of the DSMC profiles were found to have practically identical frequencies with their boundary layer counterparts. Owing to the pressure gradient in the DSMC simulations, the DSMC profiles were found to be consistently more stable than those in the model boundary layer, the differences being smaller in air than in argon. Smoothing of the DSMC profiles was found to marginally stabilize the flow. By contrast, imposition of no-slip in the boundary layer profile that modeled the DSMC data was found to have the opposite effect, which can lead to prediction of earlier transition location. However, only quantitative and no qualitative differences were found throughout the range of parameters explored. The results obtained demonstrate, for the first time, that the kinetic approach delivers flowfields that satisfy the stringent linear stability analysis requirements of accuracy of the steady basic flow and its derivatives [23, 8, 47].

Subsequently, the boundary layer developing on the flat plate was perturbed by zero net mass flux jet oscillations and two important points were demonstrated. Firstly, despite the relatively large momentum injection into the boundary layer, the unsteady perturbations generated within the DSMC framework are consistent with predictions of linear local stability theory of the underlying steady laminar profile set up when the action of the jet oscillation ceases. Secondly, the oscillations inside the boundary layer are synchronized with those generated along the leading-edge shock during the unsteady jet motion: the wavenumbers of the leading perturbation inside the boundary layer and that of the wave propagating along the shock are practically identical, while the relative difference in the frequency of the shock oscillation and that of the leading eigenmode of the steady DSMC profile is less than 4%.

In this respect, the DSMC method, which fully resolves the shock layer at all compressible flow conditions, may be optimally suited to interrogate the coupled shock layer and boundary layer regions with respect to the linear stability properties of the supersonic or hypersonic field in question.

**Acknowledgements** Access to Copper Cray XE6m has been provided by project AF-VAW10102F62 (Dr N. Bisek, PI) and is gratefully acknowledged. The authors would also like to thank the UK Turbulence Consortium and EPSRC for computational time made available on the UK supercomputing facilities ARCHER and ARCHER2 via project EP/R029326/1.

|  | Case 1<br>Air | Case 2<br>Ar | Case 3<br>$N_2$ |
|---|---|---|---|
| Mach number, $M_\infty$, [-] | 4.74 | 5.11 | 9.39 |
| Reynolds number, $Re_L$, [-] | 60354 | 71394 | 11859 |
| Prandtl number, $Pr$, [-] | 0.72 | 2/3 | 0.72 |
| Specific gas constant, $R$, [J kg$^{-1}$ K$^{-1}$] | 287.00 | 208.13 | 296.80 |
| Ratio of specific heats, $\gamma$ [-] | 7/5 | 5/3 | 7/5 |
| Plate length, $L$, [m] | 1.0 | 1.0 | 0.1 |
| Nose radius, $r_0$, [m] | $5 \times 10^{-4}$ | $5 \times 10^{-4}$ | $5 \times 10^{-4}$ |
| Free-stream velocity, $u_\infty$ [m s$^{-1}$] | 1310 | 1310 | 3000 |
| Free-stream temperature, $T_\infty$ [K] | 190 | 190 | 245 |
| Free-stream density, $\rho_\infty$ [kg m$^{-3}$] | $6.04 \times 10^{-4}$ | $8.60 \times 10^{-4}$ | $6.04 \times 10^{-4}$ |
| Free-stream viscosity, $\mu_\infty$, [N s m$^{-2}$] | $1.311 \times 10^{-5}$ | $1.578 \times 10^{-5}$ | $1.528 \times 10^{-5}$ |
| Reference viscosity, $\mu_{ref}$ [N s m$^{-2}$] | $1.719 \times 10^{-5}$ | $2.117 \times 10^{-5}$ | $1.656 \times 10^{-5}$ |
| Reference temperature, $T_{ref}$ [K] | 273 | 273 | 273 |
| Wall temperature, $T_w$ [K] | Adiabatic | Adiabatic | 245.45 |

**Table 1** Gas constants, plate geometry and pre-shock free-stream conditions

|  | Case 1<br>Air | Case 2<br>Ar | Case 3<br>$N_2$ |
|---|---|---|---|
| Knudsen number, Kn | 0.0187 | 0.0212 | 0.0897 |
| Number of Particles, Np, [-] | $1.2 \times 10^9$ | $1.2 \times 10^9$ | $7.2 \times 10^8$ |
| Number density, [Np m$^{-3}$] | $1.297 \times 10^{22}$ | $1.297 \times 10^{22}$ | $1.297 \times 10^{22}$ |
| Timestep, $dt$, [s] | $1.8 \times 10^{-8}$ | $1.8 \times 10^{-8}$ | $1.8 \times 10^{-9}$ |
| Transient period [timesteps] | 300000 | 300000 | 250000 |
| Samples | 80000 | 80000 | Instant |
| Mean free path, $\lambda$, [m] | $9.11 \times 10^{-5}$ | $8.91 \times 10^{-5}$ | $1.01 \times 10^{-4}$ |
| Power law exponent, $\bar{\omega}$, [-] | 0.75 | 0.81 | 0.74 |

**Table 2** DSMC simulation parameters

|  | Case 1 Air | Case 2 Ar | Case 3 $N_2$ |
|---|---|---|---|
| Mach number, $M_e$, [-] | 4.55 | 4.50 | 6.99 |
| $Re = \sqrt{Re_x}$, [-] | 214.0 | 213.7 | 78.1 |
| Streamwise location, $x$, [m] | 0.70 | 0.70 | 0.04 |
| Slip velocity, $u_{\text{slip}}$, [m s$^{-1}$] | 24.28 | 34.89 | 103.62 |
| Edge velocity, $u_e$, [m s$^{-1}$] | 1301.7 | 1291.5 | 2924.0 |
| Edge temperature, $T_e$, [K] | 202.90 | 238.41 | 421.00 |
| Edge density, $\rho_e$, [kg m$^{-3}$] | $6.91 \times 10^{-4}$ | $9.59 \times 10^{-4}$ | $1.19 \times 10^{-3}$ |
| Edge pressure, $p_e$, [Pa] | 40.296 | 47.537 | 148.850 |
| Edge viscosity, $\mu_e$, [N s m$^{-2}$] | $1.376 \times 10^{-5}$ | $1.896 \times 10^{-5}$ | $2.282 \times 10^{-5}$ |
| Wall temperature, $T_w$, [K] | 887.77 | 1469.66 | 416.73 |

**Table 3** Boundary layer parameters

|  | Case 1 Air | Case 2 Ar | Case 3 $N_2$ |
|---|---|---|---|
| Length scale, $L_{\text{ref}}$ [mm] | 3.269 | 3.276 | 0.511 |
| Streamwise wavenumber, $a$ [-] | 0.20 | 0.20 | 0.32 |
| Spanwise wavenumber, $\beta$ [-] | 0 | 0 | 0 |
| cBL $\omega_r$ [-] | 0.1818 | 0.1846 | - |
| DSMC $\omega_r$ [-] | 0.1824 | 0.1850 | 0.3169 |
| cBL $\omega_i$ [-] | -0.0029 | -0.0021 | - |
| DSMC $\omega_i$ [-] | -0.0039 | -0.0039 | -0.0206 |
| cBL Dimensional frequency, f [kHz] | 11.523 | 11.586 | - |
| DSMC Dimensional frequency, f [kHz] | 11.557 | 11.610 | 288.640 |
| cBL Amplification/damping rate, [1/m] | -0.91 | -0.64 | - |
| DSMC Amplification/damping rate, [1/m] | -1.21 | -1.21 | -40.32 |

**Table 4** LST parameters and results

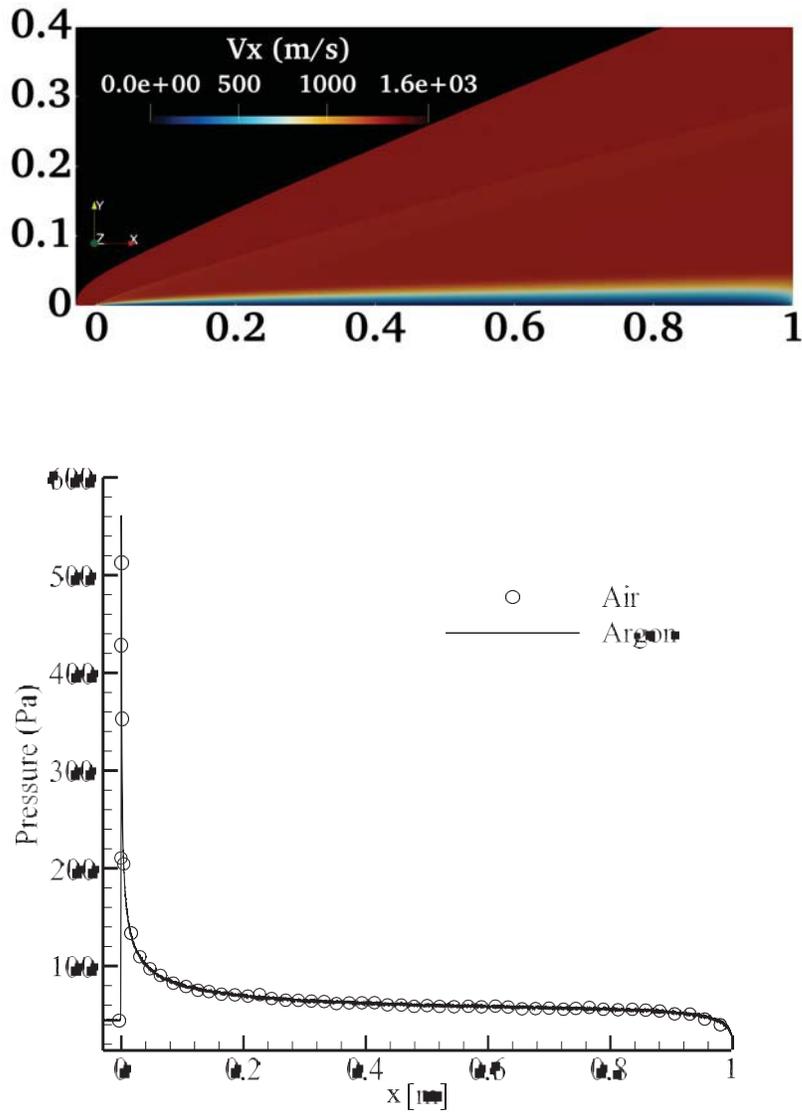

**Fig. 1** *Upper:* Streamwise velocity component at steady state in the DSMC simulation for argon. *Lower*: Pressure along the plate at the boundary layer edge ($y = 0.0014m$) for argon and air.

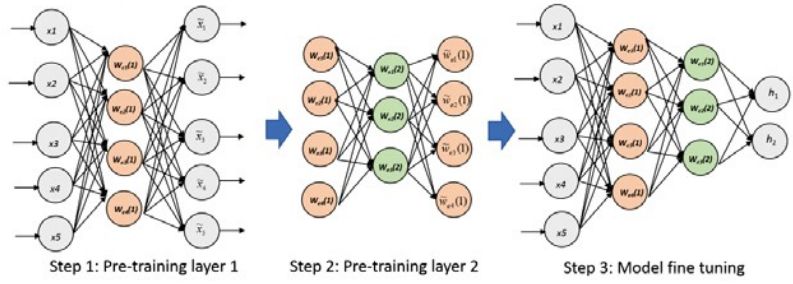

**Fig. 2** Schematic of the Stack Autoencoder Neural Network used to smooth DSMC simulation data. [42]

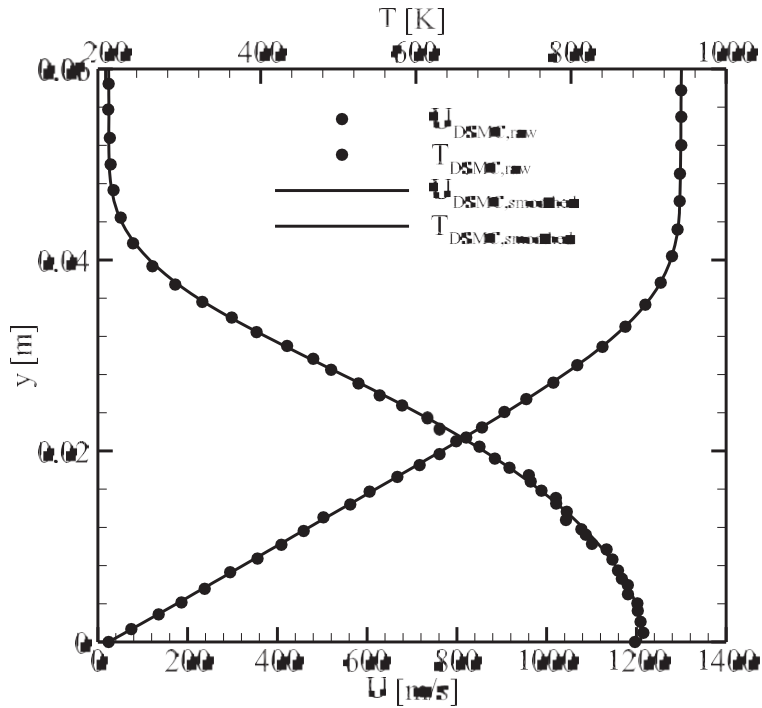

**Fig. 3** Comparison of raw DSMC data and the smoothed profiles at $x = 0.7\,m$, obtained by the neural network discussed in section 2.2 for air at the parameters of tables 1 and 2

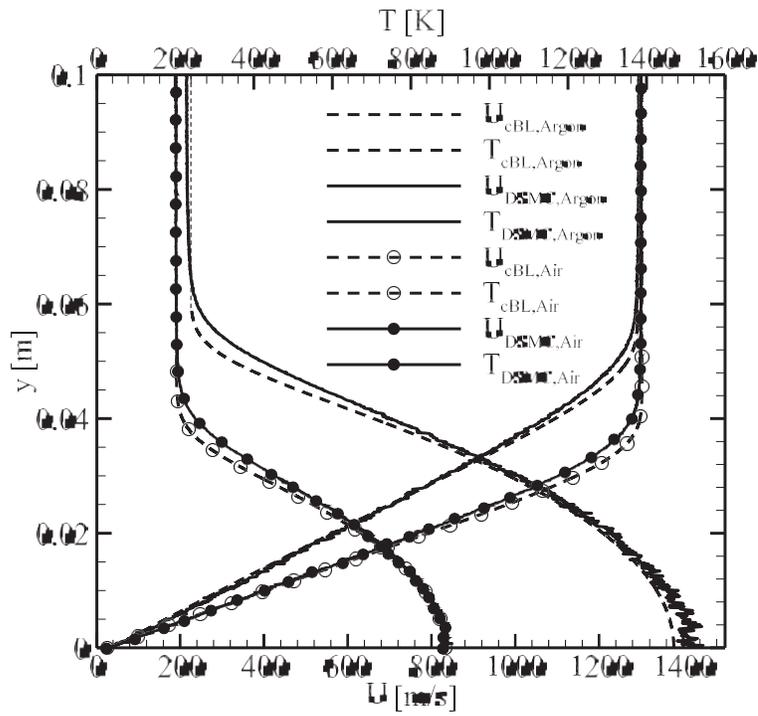

**Fig. 4** Velocity and temperature profiles obtained by DSMC in argon and air (solid) at $x = 0.7m$ and comparison with corresponding boundary layer solutions (dashed). The velocity slip computed in DSMC and calculated by equations (4-5) is shown in table 3.

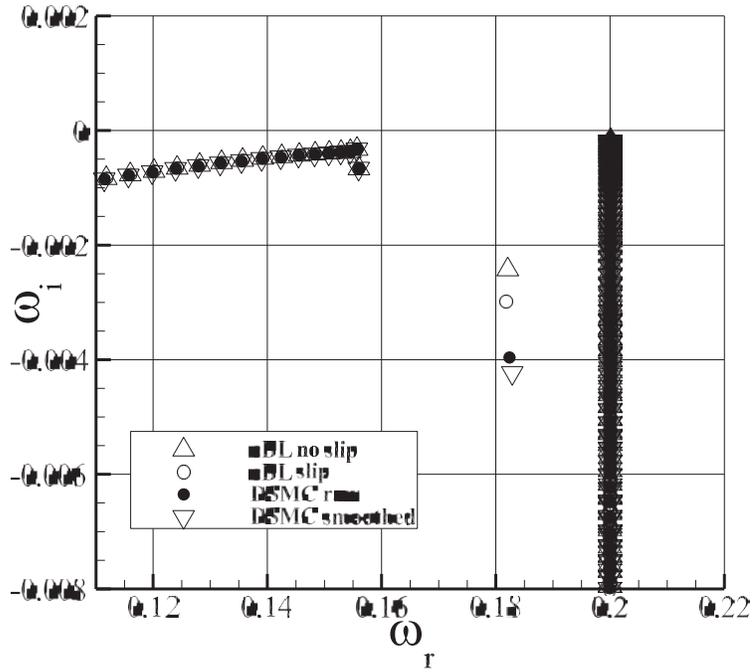

**Fig. 5** Eigenvalue spectra of raw and smoothed DSMC data, compared with compressible boundary layer (cBL) spectra under slip and no-slip boundary conditions on the base flow at the parameters of Case 1. The discrete eigenvalue obtained on the profiles including velocity slip and temperature jump for the boundary layer is $\omega_{\text{BL}} = 0.181859 - 0.00298i$, while that corresponding to the raw DSMC data is $\omega_{\text{DSMC}} = 0.182390 - 0.00396i$.

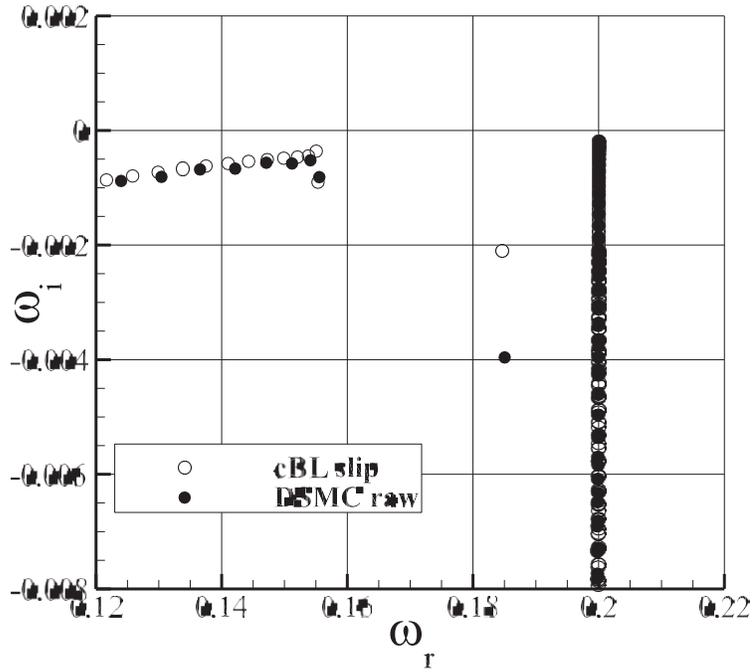

**Fig. 6** Eigenvalue spectra of DSMC and compressible boundary layer (cBL) profiles for argon at the parameters of Case 2. The respective discrete eigenvalues are $\omega_{BL} = 0.184651 - 0.00209i$ and $\omega_{DSMC} = 0.185048 - 0.00396i$.

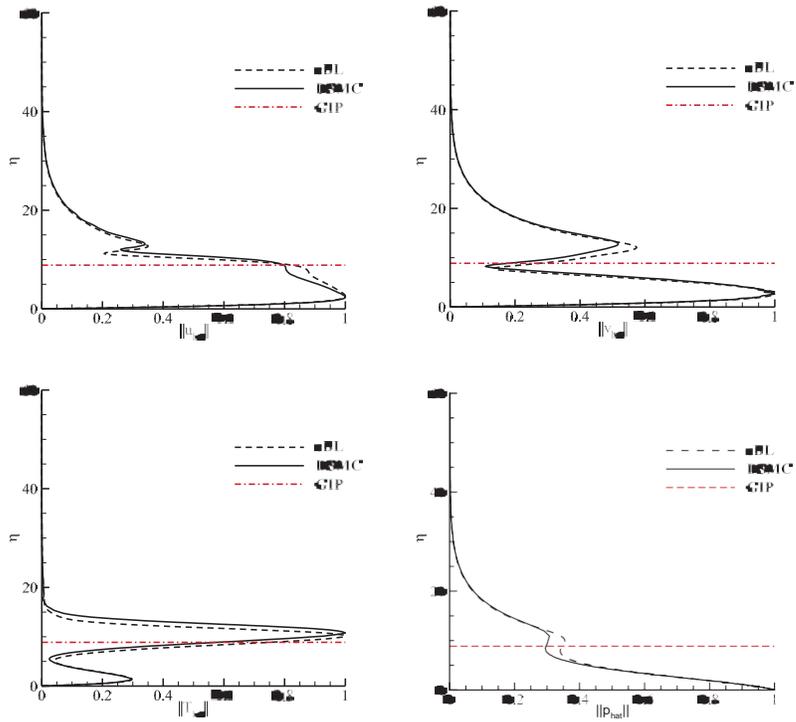

**Fig. 7** Normalized amplitude functions of linear perturbations in air, at the parameters of figure 5. The location of the generalized inflection point (GIP) is indicated by a horizontal line

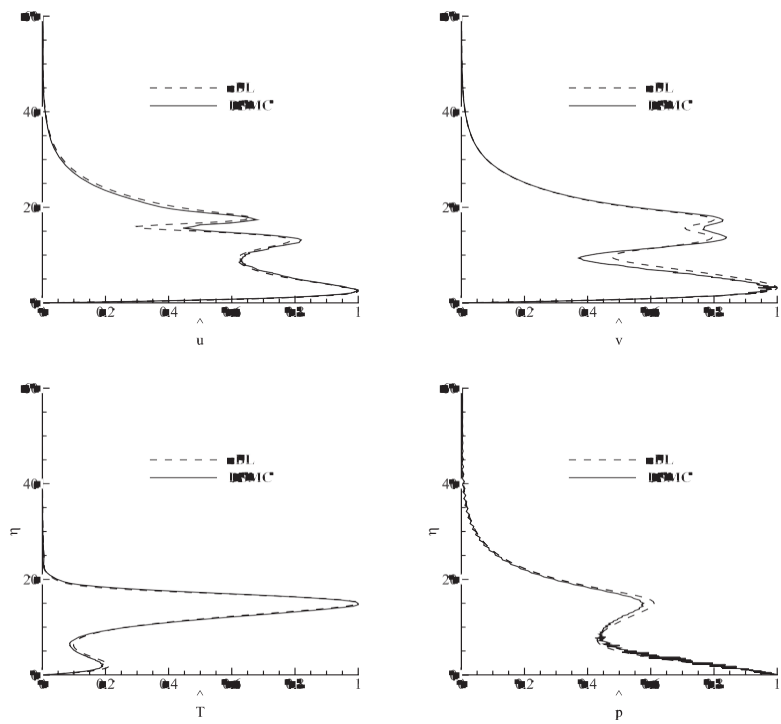

**Fig. 8** Normalized amplitude functions of linear perturbations in argon, at the parameters of figure 6.

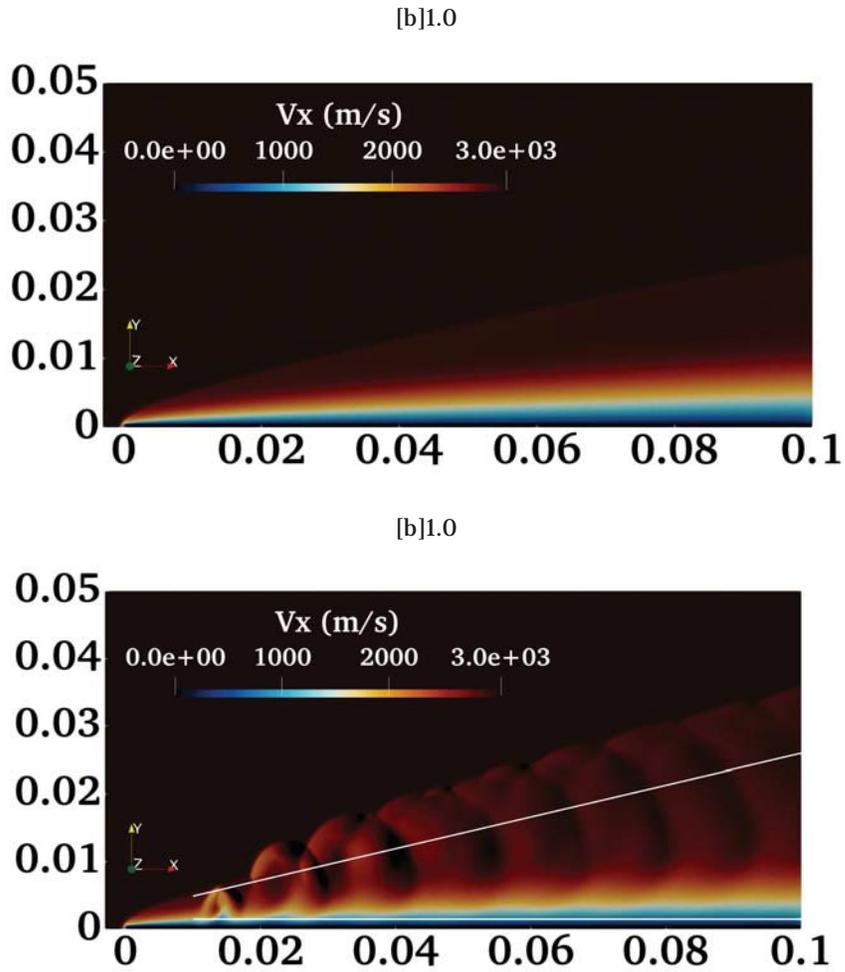

**Fig. 9** Comparison of steady state (above) and disturbed velocity field (below). The white lines show the measurements positions close to the shock and in the boundary layer. The measurements close to the shock where taken along a 14-degree angle line.

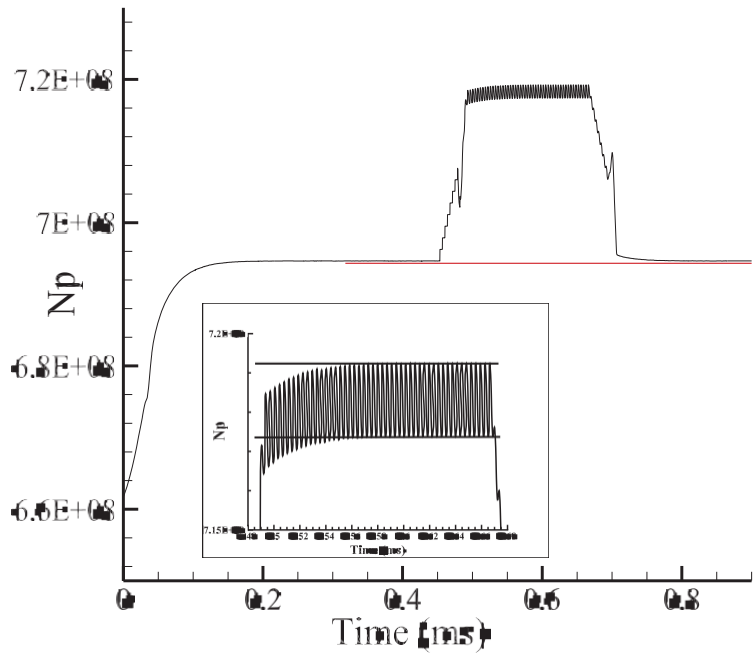

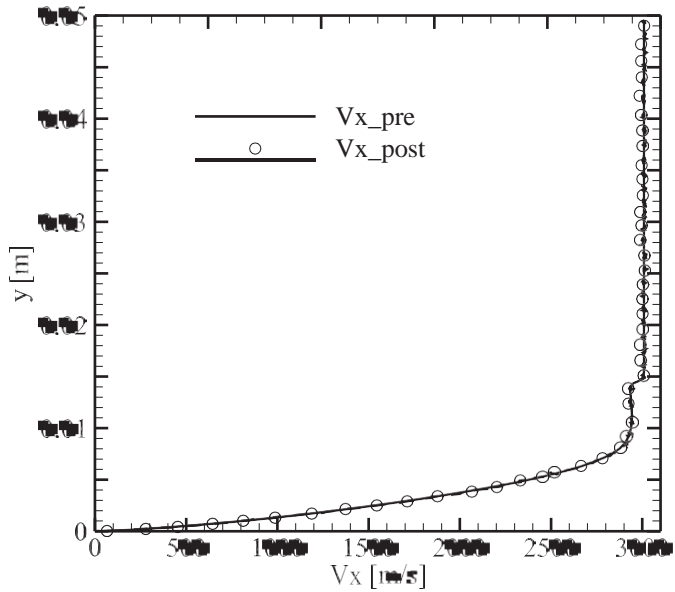

**Fig. 10** *Upper*: Total number of particles in the simulation as a function of time. The inset shows a magnified view of mass-neutral oscillations generated. *Lower:* Wall-normal velocity component, sampled at steady state at $x = 0.05m$, and instantaneous velocity obtained at $t = 0.75ms$, after the jet stops emitting; the location of the shock is clearly visible at $y \approx 0.015m$.

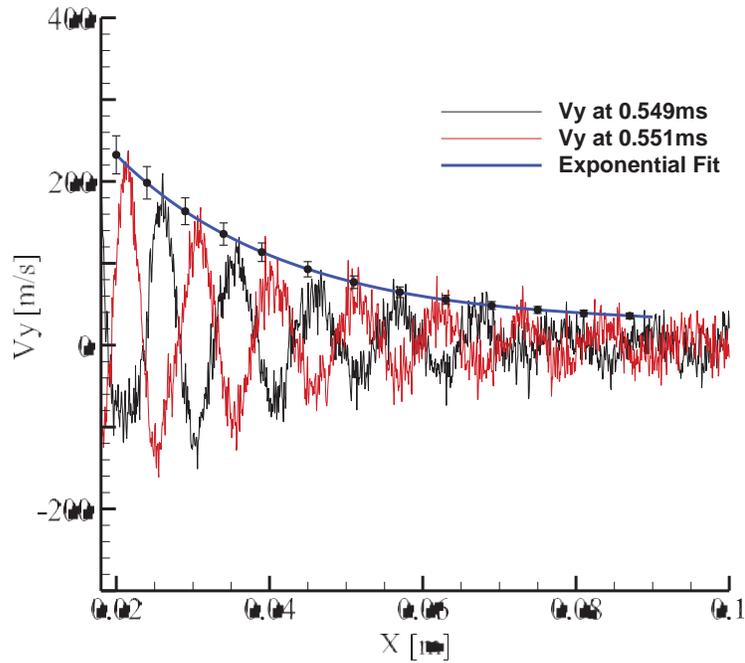

**Fig. 11** Wall-normal velocity component inside the boundary layer at a height of 0.0015m from the wall, as a function streamwise coordinate; black and red indicate raw data at two successive time steps. Also shown in an exponential curve of the form $y = Ae^{\sigma x} + C$ with $A = 508.86, \sigma = -40.32$ and $C = 25.60$. On the curve fit, a $\pm 10\%$ error bar in the damping rate $\sigma$ is also marked.

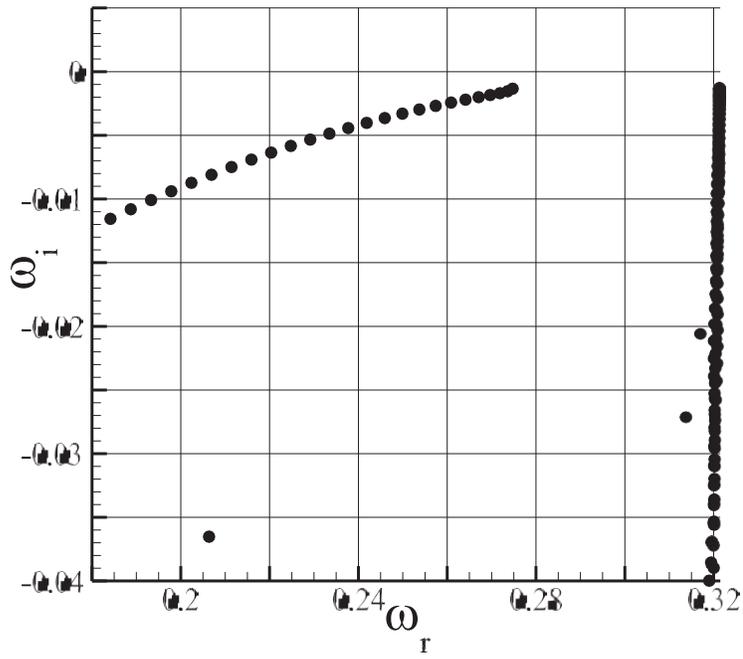

**Fig. 12** Eigenvalue spectrum of the DSMC profile of nitrogen flow at $M_e = 6.99$, $\sqrt{Re_x} = 78.1$ and wavenumbers $\alpha = 0.321$ and $\beta = 0$; quantitative results for the least-damped discrete mode are shown in table 4.

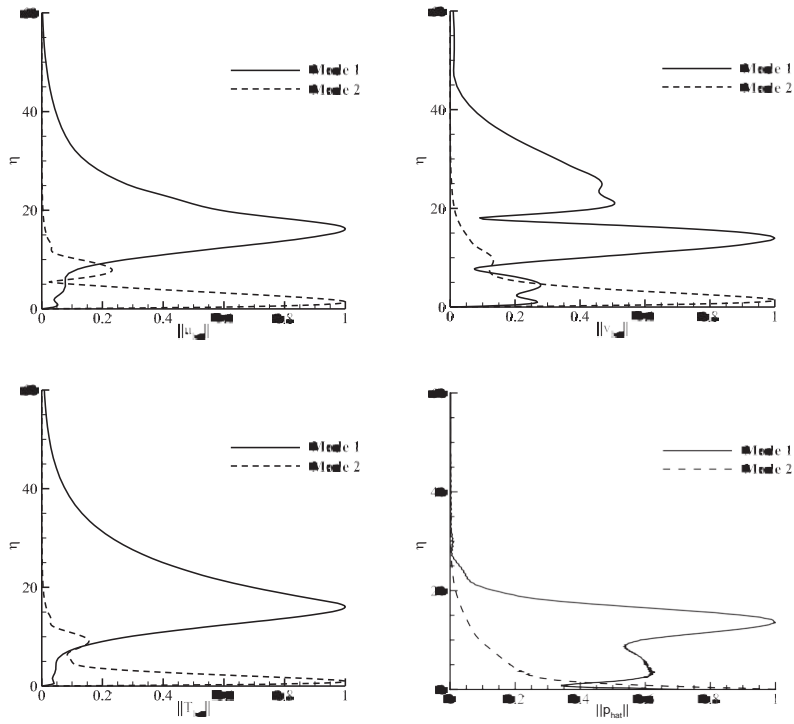

**Fig. 13** Normalized disturbance amplitude functions corresponding to the leading discrete damped mode shown in in figure 12. Shown with dashed lines are the amplitude functions of the second in significance Mode2, actually a TS wave.

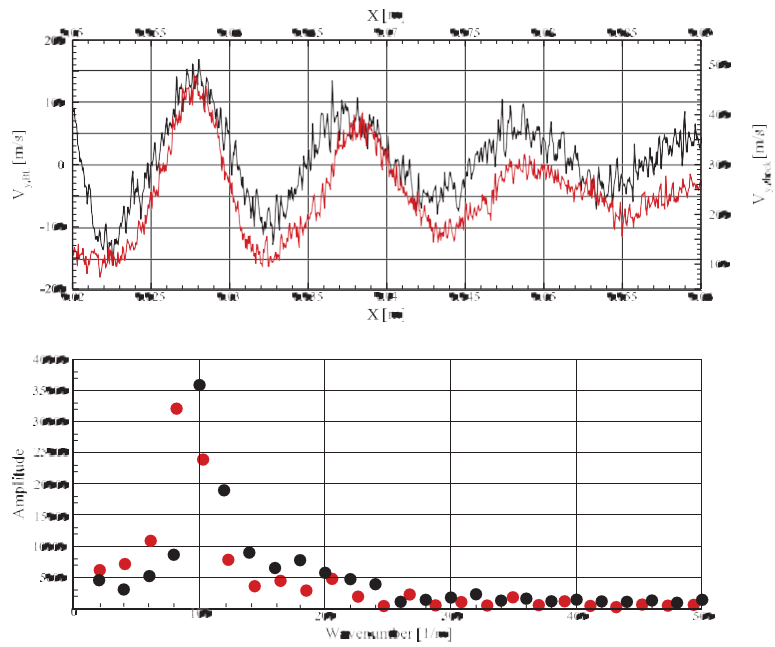

**Fig. 14** *Upper:* Streamwise disturbance amplitude as extracted from the DSMC signal in the boundary layer (black) and in the shock (red). *Lower:* Fourier decomposition of two signals using the same color code.

Linear stability analysis of hypersonic boundary layers computed by a kinetic approach     33